\documentclass[%
 reprint,
 amsmath,amssymb,
 aps,
]{revtex4-1}

\usepackage{graphicx}% Include figure files
\usepackage{dcolumn}% Align table columns on decimal point
\usepackage{bm}% bold math
\usepackage{xcolor}

\usepackage{float}

\newcommand{\Zcomm}[1]{}

\begin{document}

%\preprint{APS/123-QED}

\title{Optimizing a jump-diffusion model of a starving forager}

\author{Nikhil Krishnan}
 \email{nikhil.krishnan@colorado.edu; zpkilpat@colorado.edu }%Lines break automatically or can be forced with \\
\author{Zachary P. Kilpatrick}%
\affiliation{
    Department of Applied Mathematics, University of Colorado, Boulder, Colorado 80309, USA
}%

\begin{abstract}
We analyze the movement of a starving forager on a one-dimensional periodic lattice, where each location contains one unit of food. As the forager lands on sites with food, it consumes the food, leaving the sites empty. If the forager lands consecutively on $s$ empty sites, then it will starve. The forager has two modes of movement: it can either {\em diffuse}, by moving with equal probability to adjacent sites on the lattice, or it can {\em jump} to a uniformly randomly chosen site on the lattice. We show that the lifetime $T$ of the forager in either paradigm can be approximated by the sum of the cover time $\tau_{\rm cover}$ and the starvation time $s$, when $s$ far exceeds the number $n$ of lattice sites. Our main findings focus on the hybrid model, where the forager has a probability of either jumping or diffusing.  The lifetime of the forager varies non-monotonically according to $p_j$, the probability of jumping.  By examining a small system, analyzing a heuristic model, and using direct numerical simulation, we explore the tradeoff between jumps and diffusion, and show that the strategy that maximizes the forager lifetime is a mixture of both modes of movement.
\end{abstract}

\pacs{Valid PACS appear here}% PACS, the Physics and Astronomy
                             % Classification Scheme.
%\keywords{Suggested keywords}%Use showkeys class option if keyword
                              %display desired
\maketitle

%\tableofcontents

%Notation: $s$ steps until death, $T$ as total lifetime, $n$ as ring size, $s\leq T \leq ns$.

\section{\label{sec:intro}Introduction}
Virtually all motile organisms must forage for resources such as food, habitats, or mates.  Optimal foraging theory typically examines what strategies best balance search cost with reward~\cite{Bell90}.  An integral component of foraging is the balance between exploiting the known and/or nearby resources versus exploring one's broader environment for new resources~\cite{hills15}.  Since environmental locations have finite resources, the organism has a diminishing rate of return by remaining near its current location~\cite{stephens08} (though this can be offset if depletion is slow and resources are renewable~\cite{chupeau15}). Thus, the organism must compare the known yield at its current location with distribution of possible yields obtained by foraging distant sites~\cite{stephens87}.

The predictions of theoretical models of foraging strongly depend on how much information about the environment they assume is available to the forager. If foragers have full knowledge of the statistical distribution of resources in their environment, optimal foraging strategies are usually straightforward to identify and typically balance an explore/exploit tradeoff~\cite{Bell90,charnov76}. In contrast, foragers may possess no knowledge of their environment and may be incapable or unwilling to learn based on their foraging history~\cite{benichou14}.  Recent models along these lines study the dynamics of foragers moving in environments organized on a lattice, according to a random walk.  Previous work has examined the effect of making the forager more or less likely to pursue food~\cite{bhat17}, making the forager wait before consuming food~\cite{Benichou18}, and giving the forager a chance not to consume encountered food~\cite{rager18}. In particular, this recent work has studied the added constraint of starvation, whereby the forager cannot go longer than $s$ steps without food. Exploration/exploitation tradeoffs are then determined by how search strategy parameters shape the lifetime of the forager, corresponding to the number of steps until it starves.

Our model is similar to a starving forager executing a random walk developed in \cite{bhat17,benichou14}.  We consider the movement of a forager on a one-dimensional periodic lattice with $n$ sites, where each location contains one unit of food. If the forager lands on a site with food, the forager consumes the food, leaving the site empty. After the forager lands consecutively on $s$ empty sites, it starves.  Since the food is depleted and never regenerated, the forager will eventually starve, and can survive at most $s\cdot n$ steps, though the mean lifetime $T$ is typically much less than this upper limit.

Recent analyses have focused on cases in which foragers only move locally, according to biased or unbiased random walks. In contrast here, we explore the effects of allowing the forager to make large jumps. Food is typically distributed heterogeneously in an environment, and animals can adapt their foraging strategy as such~\cite{Bell90}.  For example, penguins alternate between foraging locally on patches of krill and moving ballistically between them~\cite{watanabe14}.  One foraging strategy for this situation is a L\`evy-type movement, where animals combine small-scale movements with long-distance displacements~\cite{lopez13,Ramos-Fernandez04,Bartumeus02,Plank08}. Our model will emulate this type of movement as follows.

\begin{figure}[t]
    \centering
    \includegraphics[scale=0.65]{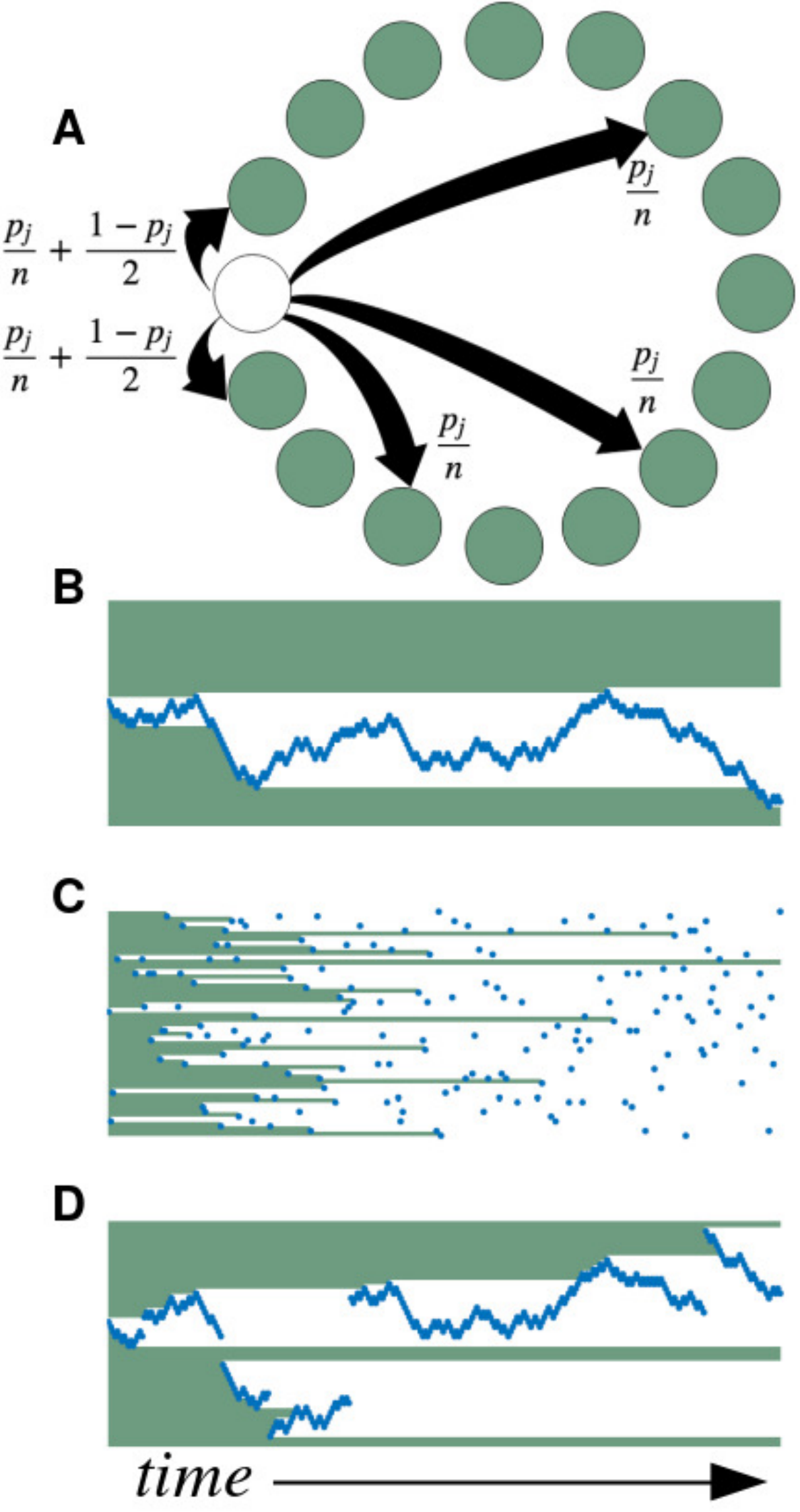}
    \caption{\Zcomm{Make the ABCD labels the same size as those in the other figures.} Jump-diffusion foraging model parametrized by $p_j$ the probability of jumping.  \textbf{A}.  The forager moves to non-adjacent sites with probability $\frac{p_j}{n}$ and to adjacent sites with probability $\frac{p_j}{n} \frac{1-p_j}{2}$, accounting for the possibility of diffusion.  \textbf{B},{\bf C},\textbf{D}. Example forager paths (blue lines/dots) for $p_j=0,1,0.03$.  Green represents sites with food, while white represents empty sites.}
    \label{fig:examplepaths}
\end{figure}

Our forager has two modes of movement: it can either diffuse, by moving with equal probability to adjacent points on the lattice  (Fig. \ref{fig:examplepaths}B), or it can jump to a uniformly randomly chosen site on the lattice  (Fig. \ref{fig:examplepaths}C). In particular, we examine a hybridized approach, where the forager jumps with probability $p_j$, or diffuses with probability $1-p_j$  (Fig. \ref{fig:examplepaths}D). Providing our forager with both types of movement allows us to consider how much time the forager should spend exploiting a given location, and how frequently the forager should move to other locations.  We demonstrate that the mean lifetime $T$ of the forager varies non-monotonically with respect to $p_j$. Specifically, we show that for non-trivial domains, the longest lifetime for the forager is obtained through a mixture of jumping and diffusion.

Obtaining an explicit formula for the forager lifetime proves difficult, perhaps even intractable. Thus, we employ a number of alternative methods for gaining insight into how the mean forager lifetime $T$ depends on model parameters.  First, we study the two boundary cases of pure diffusion and pure jumping.  In both cases, we can determine an upper bound for the forager lifetime as the sum of the cover time and survival time, and explicitly derive formulas for the forager lifetime.  This reveals that a diffusive strategy is more advantageous when the survival time $s$ is longer, whereas a jumping strategy is better for short survival times. Next, we analyze the jump-diffusion model in a very small environment (with $n=4$ food sites) and short survival time ($s=2$), showing mean lifetime is optimized by using a mix of jumping and diffusion.  Finally, we analyze a jump-wait model, where we replace the diffusive behavior with waiting behavior where the forager remains in the same location until jumping. The qualitative performance of this model is similar to the jump-diffusion model, suggesting that foragers extend their lifetime by simply not consuming food when they have recently fed.

\section{The optimal jump rate}

\begin{figure}[t]
    \centering
    \includegraphics[scale=0.25]{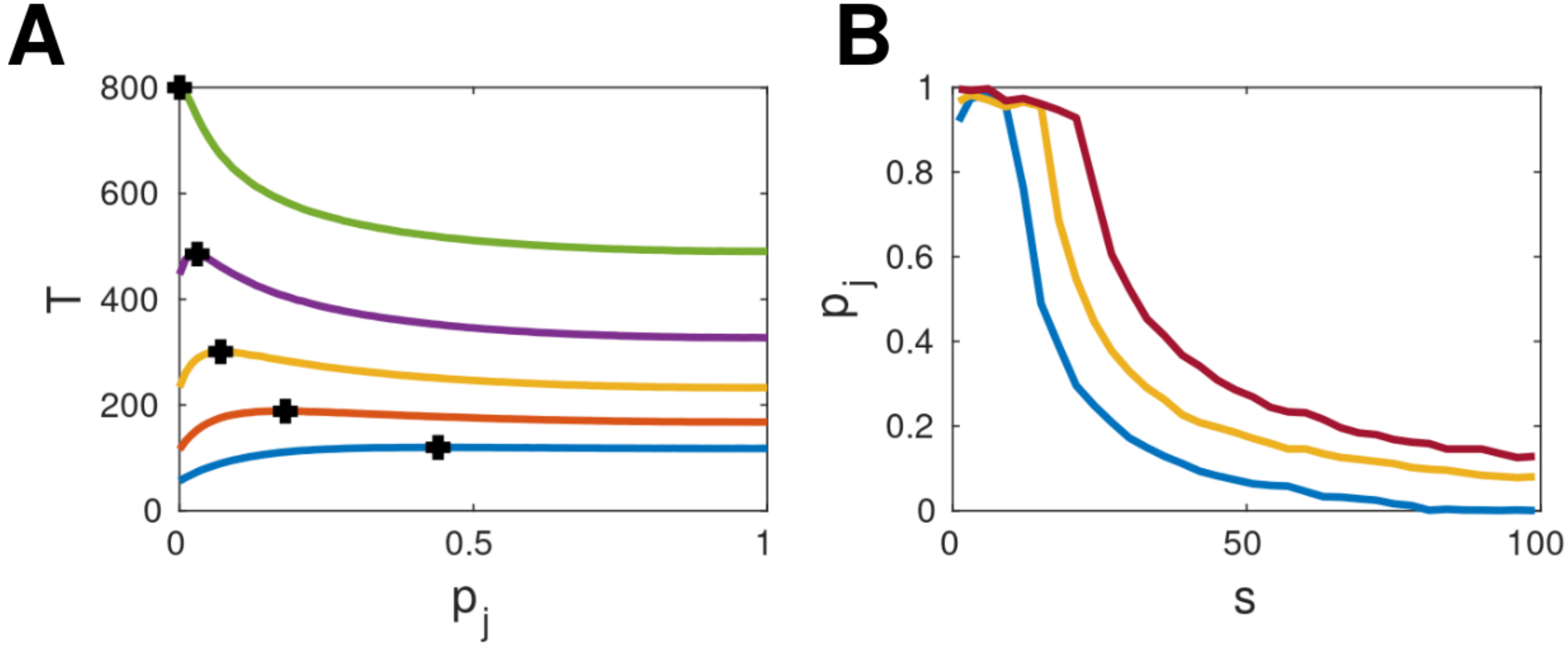}
    \caption{\textbf{A}. Mean survival time $T$ of a starving forager obeying jump-diffusion with jump rate $p_j$. Forager lifetime $T$ varied nonmonotonically with $p_j$ for $s=320,160,80,40,20$ (top-to-bottom).  The maximal lifetime is marked in black. Environment size $n=40$.  Curves are generated with $10^6$ Monte Carlo simulations. \textbf{B}. The jump rate $p_j$ that maximizes the forager lifetime primarily decreases as a function of $s$.  Shown for $n=400, 80, 20$ (top-to-bottom). Maxima are found using golden-section search~\cite{press07} using $10^6$ simulations per point.}
    \label{fig:probdiff}
\end{figure}

To begin, we consider the full hybrid model, where the forager can both jump and diffuse. We will numerically determine the effect of $p_j$ on the mean forager lifetime $T(n,s,p_j)$, while varying the environment size $n$ and survival time $s$. Across a wide range of parameters, we a mixture of jumping and diffusion ($0<p_j<1$) leads to higher values of $T$. For larger $s$ relative to $n$, the value of $p_j$ that maximizes $T$ becomes smaller. This trend will be studied in detail by analyzing related models in subsequent sections.  Numerical results are shown in Fig. (\ref{fig:probdiff}A,B). From Fig.~\ref{fig:probdiff}A, we can see $T$ is non-monotonic in $p_j$ for different values of $s$, so there is an interior $p_j$ that maximizes $T$.  As we demonstrate in subsequent sections, a larger $p_j$ (more jumping) causes the forager to consume food more rapidly, lowering the odds of starving between feedings, but depleting the resources more rapidly.  Thus, the optimal $p_j$ balances the tradeoff of slowing the rate of food consumption (decreasing $p_j$) with decreasing the probability of starving early on (increasing $p_j$). As the survival time $s$ is increased, the optimal value of $p_j$ decreases, since the forager becomes less likely to die between feedings (Fig.~\ref{fig:probdiff}B). Utilizing diffusive motion (lower $p_j$) more often limits that rate at which food is consumed. On the other hand, as the size of the environment is increased (larger $n$), the optimal $p_j$ increases. This is because there is more food initially available, so the forager can afford to increase the rate of food consumption to decrease their probability of starving.

Our interpretations of the mean lifetime $T$ dependence on $p_j$, $s$, and $n$ can be analyzed in further detail by considering a few different limiting cases and approximations of the jump-diffusion model. We begin by studying the behavior of the model at the two extremes of pure diffusion ($p_j=0$: Fig.~\ref{fig:examplepaths}B) and pure jumping ($p_j=1$: Fig.~\ref{fig:examplepaths}C). Our two main findings in this analysis are that (a) a diffusive forager covers the environment more slowly, decreasing the rate of food consumption as discussed above; and (b) jumping is a better strategy in large environments (large $n$) with lower survival times (small $s$). Indeed this is consistent with our numerical results above. We conclude with an analysis of two simpler models that demonstrate the same nonmonotonicity of $T$ in $p_j$ as shown in Fig.~\ref{fig:probdiff}.

\section{\label{sec:cover}Cover times at extremes}
Considering the boundary cases of pure diffusion $p_j=0$ and pure jumping $p_j=1$ allows us to derive explicit formulas for how model parameters, such as the environment size $n$ and starvation time $s$ impact the mean lifetime $T$ of the forager. This can be approximated first by calculating the mean cover time $\mathbb{E}(\tau_{cover})$ of the forager: the time it takes the forager to reach all of the food sites in the environment. This mean of this quantity plus the starvation time $s$ constitutes an upper bound on the lifetime in general, but for large $s$ it provides a reasonable approximation of
\begin{align} \label{cover}
    T(n,s,p_j) \approx s+\mathbb{E}(\tau_{cover}(n,p_j)).
\end{align}
This is because, when $s$ is large, the forager generally consumes almost all of the food in the domain before dying, since it will typically have enough time between feeding to locate remaining food in the environment.
%In this case, we can approximate the mean lifetime as it depends on the environment size $n$, starvation time $s$, and jump rate $p_j \in \{0,1\}$ as
%Hence, we can bound the expected lifetime of the forager from above by the expected amount of time it takes to eat the food at every site on the lattice, i.e. the cover time, plus the time it takes to starve after all food has been consumed.
%\Zcomm{This is in fact an upper bound, so maybe we can just call it an upper bound, rather than an approximation.}

The mean cover time $\mathbb{E}(\tau_{cover})$ can be computed explicitly. If $t_k$ denotes the time the $k^{th}$ piece of food is eaten, then $\tau_{cover}=t_n$, $t_1=0$, and by the linearity of expectation, we have:
\begin{align} \label{coversum}
    \mathbb{E}(\tau_{cover})=\sum_{k=2}^{n}\mathbb{E}(t_{k}-t_{k-1}).
\end{align}
In the case of both pure diffusion ($p_j=0$) and pure jumping ($p_j=1$), $\mathbb{E}(t_{k}-t_{k-1})$ can be explicitly calculated. 
\subsection{\label{sec:coverdiff}Diffusion}
We first consider the case where $p_j=0$, so the forager moves only to adjacent sites.  Following along the lines of \cite{levin08}, to calculate the cover time, we first consider the time between eating the $k^{\rm th}$ piece of food and the $k-1^{\rm th}$ piece of food.  Since the forager can only move to adjacent locations, after eating $k-1$ pieces of food, it must be on the boundary of a contiguous region of $k-1$ sites with no food -- a desert~\cite{bhat17}.  If we label the current location of the forager as site 1, and the opposite end of the desert as site $k-1$, then the time to consume the $k^{\rm th}$ piece of food is simply the hitting time of either site 0 or site $k$.  We let $f_i$ be the average time to hit either state $0$ or state $k-1$ starting at state $i$, as described by the recursion relation
\begin{align} \label{diffrecur}
    f_i&=\frac{1}{2}(f_{i-1}+1) +\frac{1}{2}(f_{i+1}+1)
\end{align}
with $f_0=f_k=0$. We can then solve Eq.~(\ref{diffrecur}) for $f_i= i(k-i)$, and note that
\begin{align*}
\mathbb{E}(t_k-t_{k-1})=f_1=k-1,
\end{align*}
so by plugging into Eqs.~(\ref{cover}) and (\ref{coversum}), we find
\begin{equation}
T(n,s,p_j=0) \approx s+\frac{n(n-1)}{2}. \label{diffcover}
\end{equation}
We can see that this approximation is linear in $s$ and quadratic in $n$, the size of the environment. Fig.~\ref{fig:ringdiff}A demonstrates that as $s$ increases, Eq.~(\ref{diffcover}) becomes more accurate, as the forager generally consumes almost all of the food in the environment. For this to be true, $s$ must be nearly an order of magnitude larger than $n$. When $s$ is too small, the forager will typically die before it can consume all of the food, so the cover time approximation breaks down.
%\Zcomm{Provide some interpretation of the above formula, basically just note it is linearly increasing in $s$ and quadratically increasing in $n$. Also, add some text referring to Fig. 2A, explaining why the approximation breaks down as $s$ is decreased.}

\subsection{\label{sec:coverjump}Jumping}
\begin{figure}[t]
    \centering
    \includegraphics[scale=0.25]{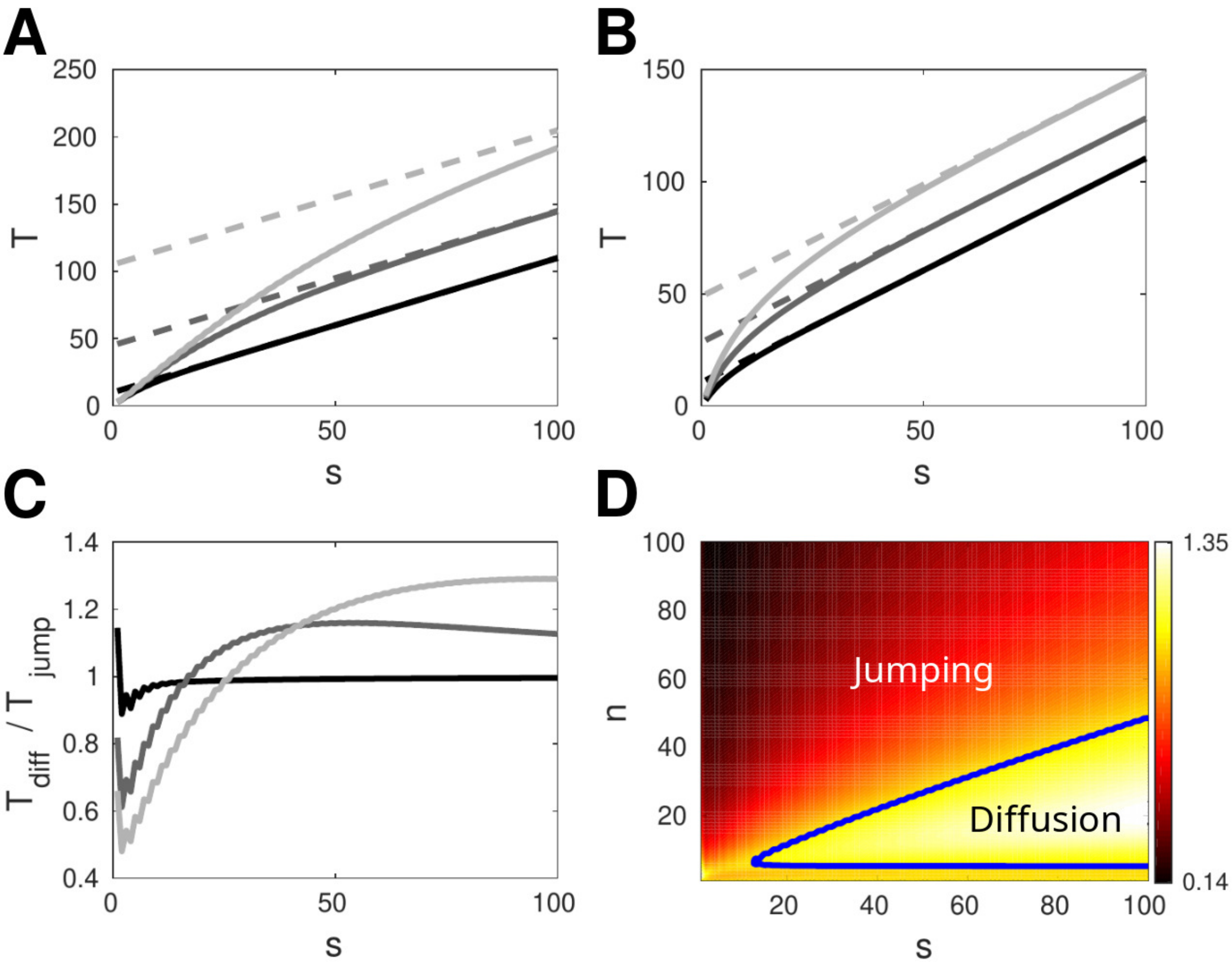}
    \caption{Forager lifetime computed from Eq.~(\ref{fullform}) in the case of pure diffusion (\textbf{A}) and pure jumping (\textbf{B}). Cover time approximations (dashed lines) computed from Eq.~(\ref{cover}) agree in the limit of large $s$: $n=15,10,5$ for black, dark grey, and grey. \textbf{C},\textbf{D}. Ratio of forager lifetime for pure diffusion to forager lifetime for pure jumping: $T_{\rm diff}/T_{\rm jump}$.  The contour on \textbf{D} marks where the ratio is one.}
    \label{fig:ringdiff}
\end{figure}

We next study the case in which the forager always jumps to a uniformly randomly chosen site on each timestep ($p_j=1$).  The cover time is then precisely the solution to the `coupon collecting problem'~\cite{levin08}.  Assume the forager has eaten $k-1$ pieces of food.  There are then $n-(k-1)$ pieces of food remaining, and the time it takes to eat the $k^{\rm th}$ piece of food is geometrically distributed:
\begin{equation*}
    (t_k-t_{k-1}) \sim \text{Geometric}\, \left( \frac{n-k+1}{n} \right).
\end{equation*}
Plugging this result into Eqs.~(\ref{cover}) and (\ref{coversum}) yields
\begin{equation}
T(n,s,p_j=1) \approx s+n\sum_{k=1}^{n-1} \frac{1}{k}. \label{jumpcover}
\end{equation}
Fig.~\ref{fig:ringdiff}B shows the exact lifetime converges to this approximation as $s$ is increased. Eq.~(\ref{jumpcover}) is again linear in $s$, but now scales much more slowly in $n$ than in the case of pure diffusion. In the limit of large $n$, we can estimate the scaling in $n$ as follows:
\begin{align*}
    n\sum_{k=1}^{n-1} \frac{1}{k}\leq n\sum_{k=1}^{n} \frac{1}{k}\leq 2n\int_{1}^n \frac{dx}{x}=2n\log(n).
\end{align*}
In particular, if we compare the cover times of the two boundary cases, we see that $\tau_{cover}(p_j=0)=\mathcal{O}(n^2)$ while $\tau_{cover}(p_j=1)=\mathcal{O}(n\:\log(n))$. This shows that if the forager can consume almost all of the food, then for large $n$, it will live longer by diffusing rather than jumping. This suggests that as $s$ increases, the optimal value of $p_j$ goes to 0, and this is indeed the case.

\section{\label{sec:ring}Forager lifetime at extremes}
We now determine the exact formula for the lifetime of the forager. While the formula we derive actually applies to all values of $p_j \in [0,1]$, we can only compute its constituent parts explicitly in the boundary cases $p_j \in \{0,1\}$.  Let $X_k$ denote the time between eating the $k^{th}$ piece of food and the $k-1^{th}$ piece of food, where $X_1=0$, since the forager immediately consumes food at their initial position. The probability that the forager, with starvation time $s$, consumes $k$ pieces of food thus equals
\begin{align}
\mathbb{P} (k^* = k) = \mathbb{P}(X_1,...,X_k\leq s,X_{k+1}>s),  \label{Pks}
\end{align}
so $k^* \in \{1,...,n\}$ is a random variable arising from the stochastic movement and death of the forager. We can determine the distribution over $k^*$ by first computing the cumulative distribution for each $X_k$:
\begin{align}
    &F_k(s)=\mathbb{P}(X_k \leq s)=\sum_{j=1}^s \mathbb{P}(X_k=j), \: F_{n+1}(s)=0.  \label{Xkcum}
\end{align}
Thus, $F_k(s)$ is the probability the forager survives long enough to consume the $k^{th}$ piece of food having consumed the $k-1^{th}$ piece of food.  We also wish to know the probability that the forager eats exactly $k$ pieces of food before dying.  This is given by
\begin{align*}
\mathbb{P}(k^* = k)= \Big(1-F_{k+1}(s)\Big)\prod_{j=1}^k \Big(F_j(s)\Big).
\end{align*}
The forager lifetime in each case can be computed first by conditioning on consuming exactly $k$ pieces of food, $T | k^* = k$, which is simply the time it takes to eat the $k$ pieces of food, plus $s$ steps more until starvation. The expected lifetime $T$ is then given by marginalizing over all possible values of $k^*$:
\begin{align} \label{fullform}
    &\mathbb{E}(T)= \left[ T | k^* = k \right] \mathbb{P} (k^* = k ) \\
    &=s+\sum_{k=1}^{n}\Big(1-F_{k+1}(s)\Big)\prod_{j=1}^k F_j(s) \sum_{i=1}^k \mathbb{E}(X_i |X_i \leq s). \notag
\end{align}
For the boundary cases of $p_j=0$ and $p_j=1$, we can derive an explicit formula for $\mathbb{P}(X_k=j)$ in Eq.~(\ref{Xkcum}), allowing us to explicitly calculate Eq.~(\ref{fullform}).  In the limit of large $s$, we can approximate $F_j(s) =1$ for all $j \leq n$ and reduce Eq.~(\ref{fullform}) to the sum of the expectations, $\mathbb{E}(X_i)$, which is the cover time upper bound given in Eq.~(\ref{cover}).
%\Zcomm{At this point, I think it would help to also note that in the limit of large $s$, we can approximate $F_j(s) =1$ for all $j \leq n$. Therefore the above formula simply reduces to the sum of the expectations, $E[X_i]$, which is the cover time upper bound derived above. N: Almost verbatim, seemed like the most concise way to state it}

\subsection{\label{sec:ringdiff} Diffusion}
When $p_j = 0$, the forager moves by diffusion to carve out a food desert, a simply connected region without any food.  As before, we label the sites of the desert so that the first boundary of the desert is site $1$ while the last boundary is site $k-1$.  Following \cite{duyzend08,feller68}, we can determine the probability mass function of $X_k$.

Let $u_{\ell,j}$ be the probability that it takes exactly $j$ steps to first hit site 0 from site $\ell$.  We then have the following recursion relation:
\begin{align} \label{initrec}
    &u_{\ell,j+1}=\frac{1}{2}u_{\ell-1,j}+\frac{1}{2}u_{\ell+1,j}\\
    &u_{0,0}=1,\:u_{j,0}=u_{0,j}=u_{k,j}=0,\forall j>0. \notag
\end{align}
We then define the generating function
\begin{equation*}
    U_\ell(v)=\sum_{j=0}^\infty u_{\ell,j} v^j
\end{equation*}
and multiply Eq.~(\ref{initrec}) by $v^{j+1}$, so that by summing over $j$ we obtain
\begin{align} \label{medrec}
    &U_\ell(v)=\frac{v}{2}U_{\ell-1}(v)+\frac{v}{2}U_{\ell+1}(v)\\
    &U_0(v)=1,\: U_k(v)=0. \notag 
\end{align}
The boundary conditions arise from the fact that the probability of hitting site 0 is $u_{0,0}=1$ if starting there, but $u_{k,0} = 0$ if starting at the opposite food site. Considering solutions to Eq.~(\ref{medrec}) of the form $U_\ell(v)=\lambda^\ell(v)$, we obtain the characteristic equation
\begin{align*}
    \lambda(v)=\frac{v}{2}+\frac{v}{2}\lambda^2(v).
\end{align*}
This quadratic equation has two roots:
\begin{align} \label{quadroot}
    \lambda_{\pm}(v)=\frac{1\pm \sqrt{1-v^2}}{v},
\end{align}
assuming $0<v<1$.  Each root is a particular solution to Eq.~(\ref{medrec}).  It follows that there are some functions $A(v),B(v)$ so the general solution has the form
\begin{equation}
    U_\ell(v)=A(v)\lambda_+^\ell(v)+B(v)\lambda_-^\ell(v).
\end{equation}
We can apply the boundary conditions $A(v)+B(v)=1$ and $A(v)\lambda_+^k(v)+B(v)\lambda_-^k(v)=0$ from Eq.~(\ref{medrec}) to determine $A(v)$ and $B(v)$. Finally, noting that by their definition, $\lambda_+(v)\lambda_-(v)=1$, we have the explicit formula:
\begin{equation} \label{generator}
    U_\ell(v)=\frac{\lambda_+^{k-\ell}(v)-\lambda_-^{k-\ell}(v)}{ \lambda_+^k(v)-\lambda_-^k(v)}.
\end{equation}
The quantity $U_\ell(v)$ is the quotient of polynomials.  To determine $u_{\ell,j}$, we will decompose $U_\ell(v)$ with partial fractions.  To start, we make the change of variables $v=\sec \phi$.  Applying this to Eq.~(\ref{quadroot}) and Eq.~(\ref{generator}), we find
\begin{align*}
    \lambda_{\pm}(v)=\cos \phi \pm i\sin \phi,  \hspace{3mm} U_\ell(v)=\frac{\sin (k-\ell)\phi}{\sin k\phi}.
\end{align*}
We can see that the denominator of $U_\ell(v)$ is zero for $\phi_m=\frac{m \pi}{k},\: m=0,...k$, which correspond to $v_m=\sec \phi_m$.  Furthermore, since the degree of the numerator exceeds the degree of the denominator by at most 1, $U_\ell(v)$ has a partial fraction decomposition with the form
\begin{equation} \label{partfrac}
    \frac{\sin (k-\ell)\phi}{\sin k\phi}=Av+B+\frac{\rho_1}{v_1-v}+...+\frac{\rho_{k-1}}{v_{k-1}-v}.
\end{equation}
To determine the value of $\rho_m$, we multiply both sides by $v_m-v$, then take $v \rightarrow v_m$ (and $\phi \rightarrow \phi_m$):  
\begin{equation*}
    \rho_m=\frac{\sin \frac{\ell \pi m}{k}\sin \frac{\pi m}{k}}{k \cos^2\frac{\pi m}{k}}.
\end{equation*}
By decomposing each fraction of Eq.~(\ref{partfrac}) into a geometric series, we can see that the coefficient of $v^j$ (which is $u_{\ell,j}$) is given by
\begin{equation*}
    \sum_{m=1}^{k-1} \frac{\rho_m}{v_m^{j+1}}=\frac{1}{k}\sum_{m=1}^{k-1} \cos^{j-1}\left( \frac{\pi m}{k} \right)  \sin \left( \frac{\ell \pi m}{k} \right) \sin \left( \frac{\pi m}{k} \right).
\end{equation*}
We are interested in two possibilities.  Either, the forager can start at site 1 and hit site 0 (corresponding to $u_{1,j}$), or the forager can start at site 1 and hit site $k$, which by symmetry is identical to the forager starting at site $k-1$ and hitting site 0 (corresponding to $u_{k-1,j}$).  The probability $\mathbb{P}(X_k=j)$ that it will take $j$ steps from consuming the $(k-1)^{th}$ to the $k^{th}$ food site is then the weighted sum of these two possibilities. Thus, we have the following distribution for $X_k$:
\begin{align*}
    \mathbb{P}(X_{k}=j)=&\frac{1}{k}\sum_{m=1}^{k-1} \cos^{j-1}\Big(\frac{\pi m}{k}\Big)\sin\Big(\frac{\pi m}{k}\Big)\\
    &\times \bigg(\sin\Big(\frac{\pi m}{k}\Big)+\sin\Big(\frac{\pi (k-1) m}{k}\Big)\bigg). \notag
\end{align*}
We can compute the corresponding conditional expectations and cumulative distributions in the standard way, and then use Eq.~(\ref{fullform}) to compute the expected lifetime of the forager.  From Fig.~\ref{fig:ringdiff}A we can see that for small values of $s$ the forager lifetime is initially super-linear in $s$, but that as $s$ increases, the lifetime slowly converges to a linear function of $s$, as described by the cover time approximation. Furthermore, the lifetime $T(n,s, p_j=0)$ is generally insensitive to $n$ for small values of $s$. This is because the forager will rarely ever consume all the food in its environment in these cases. \\

\subsection{\label{sec:ringjump} Jumping}
For $p_j=1$, the time $X_k$ between consuming the $(k-1)^{th}$ and $k^{th}$ food site is geometrically distributed with success probability $\left[n-(k-1) \right]/n$.  Specifically, $\mathbb{P}(X_k = j)$ is the probability of $j-1$ visits to empty sites, each with probability $\left[ k-1\right] / n$, followed by a visit to a food site, with probability $\left[ n-(k-1)\right]/n$.  Thus,
\begin{equation*}
    \mathbb{P}(X_k=j)=\left(\frac{k-1}{n} \right)^{j-1} \left( \frac{n-(k-1)}{n} \right).
\end{equation*}
We can compute the cumulative distributions and conditional expectations of a geometric random variable in the typical way, to yield the following formula for the forager lifetime from Eq.~(\ref{fullform}):
\begin{align*}
   T(n,s,p_j=1) &=s+\sum_{k=1}^{n-1}\bigg(\frac{k+1}{n}\bigg)^s\prod_{j=1}^k \left( 1-\bigg(\frac{j}{n}\bigg)^s \right) \\
    &\times \sum_{i=1}^k \bigg[ \frac{n}{n-i}+s+\frac{s}{(i/n)^s-1} \bigg]. 
\end{align*}
From Fig.~\ref{fig:ringdiff}B we again see that for small values of $s$ the forager lifetime $T(n,s, p_j=1)$ is initially super-linear in $s$ and insensitive to $n$, but limits to the cover time approximation as $s$ increases.

To compare the two strategies (pure diffusion vs. pure jumping), we compute the ratio $T_{\rm diff}/T_{\rm jump}$ of the forager lifetime for $p_j=0$ to the forager lifetime for $p_j=1$.  From Fig~\ref{fig:ringdiff}C, we can see several notable features.  Most importantly, for sufficiently large $s$, diffusion leads to longer lifetimes than jumping. This is because the diffusive forager will cover the environment more slowly than the jumping forager, so they will not consume food as quickly. Note, the ratio drops from $s=1$ to $s=2$, since in the case of pure diffusion (and $s=1$) the forager will live at least two time steps, whereas the pure jumper may not. As soon as $s=2$, this effect becomes negligible. Furthermore, this drop in the ratio becomes less severe for larger values of $n$, since the jumper will almost always live at least two timesteps. In Fig.~\ref{fig:ringdiff}D, we display the ratio as a surface plot along both the $s$ and $n$ axis. Increasing $s$ clearly expands the region (outlined) of $n$ values, for which diffusion is a better strategy. Note that for very small values of $n$ ($n\leq 5$) the cover time for diffusion is less than the cover time for jumping, leading to an advantage of jumping over diffusion at those parameter values.  When $s$ is large relative to $n$, the diffusive forager benefits from a larger cover time, so for these small values of $n$, it is consistently more beneficial to jump rather than diffuse. On the other hand, when $s$ is small compared to $n$, it is better to jump since this will decrease the likelihood of starving before much of the environment's food has been consumed.
%\Zcomm{You need an in text explanation of Fig. 2. In particular, explain why we are computing the ratios of $T$ for diffusion vs. jumping. It is because we want to show the ranges of values of $s$ for which jumping vs. diffusion is better. Also, can you look into why things get better in the case of diffusion at very low values of $n$? This seems counterintuitive to our general claim. N: This seems a more appropriate place to explain the figure, so your comment has been relocated.}

This concludes our analysis in the case of pure diffusion ($p_j=0$) or pure jumping ($p_j=1$). We now turn to two simpler instantiations of the jump-diffusion model of the starving forager: one that considers a very small environment ($n=4$) and another that considers replacing diffusion with waiting. Both of these models exhibit the same nonmonotonicity of the lifetime $T$ with respect to $p_j$, and admit some explicit analysis.

\section{\label{sec:probdiff}Tractable Models and Approximations of Jump-Diffusion}

Given our insights from the extreme cases $p_j \in \{0,1\}$, we now consider the full hybrid model, where the forager can both jump and diffuse.  We have seen that when $s$ is large is relative to $n$, it is more advantageous to diffuse rather than jump. To obtain explicit expressions of this result, we will consider two simplifications. First, for a sufficiently small system (small $n$ and $s$), the forager lifetime can be explicitly determined either by combinatorial methods or by analyzing the probability transition matrix for the system. Secondly, we will consider a model that replaces diffusion with waiting.  This altered model still yields qualitatively similar results to the jump-diffusion model, lending credence to our theory that diffusion acts as a way to prevent premature resource depletion. Both of these models demonstrate that it is most beneficial for the forager to use a mixture of jumping and diffusing (or waiting), specifically that $T(n,s,p_j)$ has an interior maximum on $p_j \in [0,1]$.  Furthermore, in the case of the jump-wait model, we will see that the larger $s$ is relative to $n$, the smaller the optimal value of $p_j$ becomes, consistent with our results for the jump-diffusion model.

\subsection{\label{sec:small}Small System}
\begin{figure}[t]
    \centering
    \includegraphics[width=8.5cm]{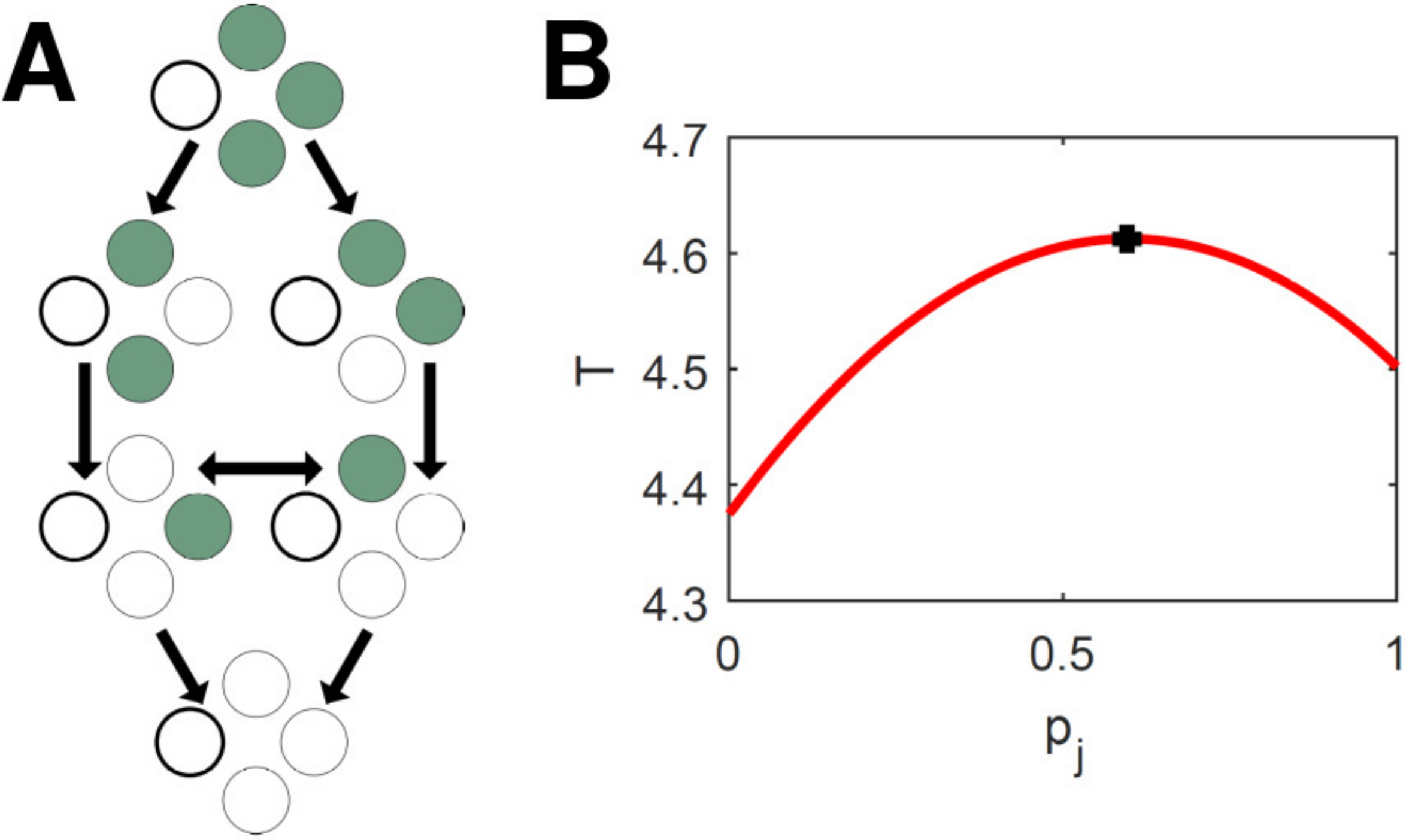}
    \caption{\Zcomm{The top of panel A appears to be cut off} Jump-diffusion model of foraging for small environment size and survival time ($n=4$, $s=2$).  \textbf{A} Enumeration of food/forager geometries for $n=4$ and $s=2$ system.  The forager is always in the left (bold) site. The arrows denote admissible transitions between geometries.  Note that since $s=2$, every geometry can return to itself once.  \textbf{B}  Expected forager lifetime $T$ as a function of $p_j$ has an interior maximum (black dot).}
    \label{fig:smallsys}
\end{figure}

For a system of small enough size, the lifetime of the forager can be analytically determined, either by enumerating all possible outcomes or by analyzing an associated discrete-time Markov chain.  Here, we consider a lattice with $n=4$ sites, and a starvation time of $s=2$.  The combination of food and forager states can be described as a thirteen state Markov chain (Fig.~\ref{fig:smallsys}A). Note that the forager can transition from most state geometries to death, by landing on a site without food more than $s=2$ times in a row. The nonzero entries of the associated transition matrix $Q$ corresponding to the probabilities to transition from state $i$ to $j$ are given in the Appendix \ref{appss}. 

To calculate the forager lifetime, we compute the mean absorption time into the thirteenth state (the cemetery state) as a passage time problem for Markov chains~\cite{norris97}.  Let us denote by $v$ be the vector of all zeroes save the first entry which is one.  Let $\textbf{1}$ be a vector of all ones.  Finally, let $Q$ be the matrix consisting of the first twelve rows and columns of the preceding probability transition matrix (the sub-matrix excluding the cemetery state).  The expected forager lifetime is then given by
\begin{equation*}
    T(4,2,p_j) =\textbf{1}^T(I-Q^T)^{-1}v.
\end{equation*}
We can also determine the expected forager lifetime $T(4,2,p_j)$ by enumerating outcomes directly, yielding the following polynomial:
\begin{align*}
    T(4,2,p_j)=&-\frac{3}{512}p_j^6-\frac{3}{256}p_j^5+\frac{15}{256}p_j^4\\
    & -\frac{11}{128}p_j^3-\frac{39}{64}p_j^2+\frac{25}{32}p_j+\frac{35}{8},
\end{align*}
which can be maximized by standard methods (Fig.~\ref{fig:smallsys}B). With either method of computation, the maximal forager lifetime is $T^{\rm max} \approx 4.612$ at $p_j \approx 0.598$, demonstrating it is optimal for the forager to both jump and diffuse in this simple case. Examining Fig.~\ref{fig:smallsys}A, we expect that the forager lifetime is lengthened by allowing the system to dwell in the intermediate states preceding the bottom cemetery state. This is the same intuition as in the large system: the optimal forager balances a reduction in their probability of starving before eating all the food with a reduction in the rate at which food is consumed.

\begin{figure}[t]
    \centering
    \includegraphics[scale=0.25]{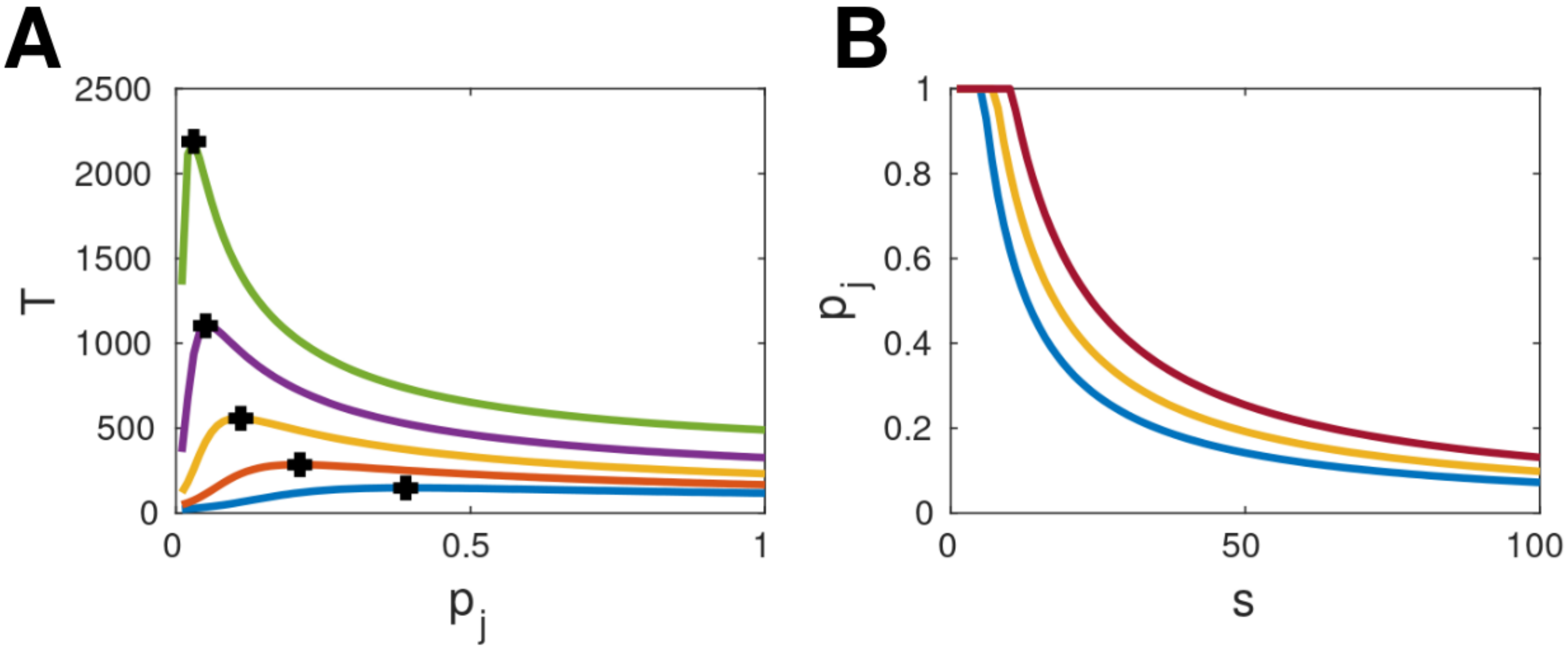}
    \caption{Jump-wait model of foraging.  \textbf{A} Forager lifetime as a function of $p_j$. The maximal lifetime is marked in black: $s=320,160,80,40,20$ top-to-bottom as in Fig.~\ref{fig:probdiff}A.  \textbf{B} The value of $p_j$ that maximizes the forager lifetime as a function of $s$: $n=400,80,20$ top-to-bottom as in Fig.~\ref{fig:probdiff}B.}
    \label{fig:probwait}
\end{figure}

\subsection{\label{sec:wait}A Jump-Wait Model}
We now consider a modification of our jump-diffusion model which admits explicit analysis as a function of the jumping probability $p_j$.  In this case, rather the forager moving to adjacent sites with probability $1-p_j$, they remain at the same site with probability $1-p_j$.  Replacing diffusion mimics diffusion in food `deserts,' which will generally arise in large domains when $p_j$ is not too large. This modified model essentially removes the food consumption and starvation-resetting that arises from diffusion and only allows this to occur when the forager jumps. As we saw in our discussion of cover times, a diffusing forager covers the domain more slowly than a jumping forager. Waiting, just like diffusing, acts to slow the rate at which the forager consumes the food, providing qualitatively similar non-monotonic lifetimes in $p_j$ (Fig.~\ref{fig:probwait}A). 

In this case, we can obtain an analytic expression for the forager lifetime, by noting that the inter-feeding times $X_k$ is geometrically distributed with success probability success probability $p_j\frac{n-(k-1)}{n}$. This can be derived by noting that the probability that the forager lands on a piece of food is the probability that the forager jumps at all, $p_j$, multiplied by the probability that the forager lands on a site with food, $\frac{n-(k-1)}{n}$.  The probability that $X_k=j$ is thus the probability of $j-1$ visits to empty sites multiplied by the probability of a visit to a site with food:
\begin{align*}
    \mathbb{P}(X_k=j)=\left(1- p_j\frac{n-(k-1)}{n} \right)^{j-1}\left( p_j\frac{n-(k-1)}{n} \right).
\end{align*}
The conditional expectation and cumulative distributions for a geometric random variable can be computed in the standard way, giving the forager lifetime from Eq.~(\ref{fullform}):
\begin{align}
    \mathbb{E}(T)=&s+\sum_{k=1}^{n-1}\Bigg[ \bigg(\frac{p_j(k-n+1)+n}{n}\bigg)^s\\
    &\prod_{\ell=1}^k (1-\bigg(\frac{p_j(\ell-n)+n}{n}\bigg)^s) \notag \\
    &\sum_{i=1}^k \bigg[ s+\frac{n}{(n-i)p_j}+\frac{s}{(\frac{p_j(i-n)+n}{n})^s-1} \bigg] \Bigg] \notag
\end{align}
Taking a large $s$ limit of this expression, we obtain:
\begin{align}
    \mathbb{E}(T)\approx s+\frac{n}{p_j}\sum_{k=1}^{n-1} \frac{1}{k},
\end{align}
which is exactly the approximation Eq.~(\ref{cover}) for the mean cover time plus the starvation time $s$. The cover time is equal to that from the case of pure jumping, Eq.~(\ref{diffcover}), scaled by $\frac{1}{p_j}$.  This demonstrates that for sufficiently large $s$, the smaller the value of $p_j$, the longer the expected forager lifetime.

By examining Fig.~\ref{fig:probwait}A, we see that the jump-wait model shares important characteristics with the jump-diffusion model.  The forager lifetime is non-monotonic in $p_j$ and the optimal value of $p_j$ decreases as $s$ increases.  Additionally, the optimal $p_j$ decreases as a function of $s$, but increases as a function of $n$ (Fig.~\ref{fig:probwait}B).  Similar to the optimal $p_j$ curves for the jump-diffusion model, the curves have sections of relatively rapid change for intermediate values of $s$. Thus, our finding for the jump-wait model again suggest that a starving forager will can maximize their lifetime by balancing a decrease in the rate of food consumption (by lowering $p_j$) with an increase probability of surviving until most of the food is consumed (by increasing $p_j$).

\section{Discussion}
We have extended the recently-developed starving forager model~\cite{benichou14} to account for the possibility of long-range motion via jumping.  The combination of these two modes of movement is related to L\`evy-type motion often found in the dynamics of motile organisms' foraging strategies~\cite{lopez13}.  By analyzing cover times, we have shown that jumping consumes food more rapidly than diffusion.  This provides an explanation for why a mixture of jumping and diffusion is optimal: excessive jumping leads to rapid food depletion, excessive diffusion leads to earlier starvation of the forager who gets stuck in food `deserts.'  This explanation is further validated by the qualitative similarities of the jump-diffusion and jump-wait models.  For each, larger $s$ or smaller $n$ each correspond to more diffusion (waiting) being optimal, while the converse corresponds to more jumping being optimal.
%\Zcom{Once the rest of the paper has been modified, I think we should update the Discussion. N: ?}

Our model of a starving forager with a mixture of movement modes suggests several possible extensions.  Throughout this work, jumping has represented movement with equal probability to any lattice site.  However, a forager executing a jump may more often select a site that is further away, to avoid revisiting empty sites. They may also be less likely to make extremely large jumps. This would suggest a model where the jump process is associated with a nonuniform distribution of jump distances. Also, our work has only considered a periodic one-dimensional lattice environment.  The behavior of the forager in higher dimensions is still open, and it would be interesting to see how the forager lifetime depends on domain size and geometry in higher dimensions (e.g., plane, torus, or sphere).  Another relevant extension would be for the forager to retain some information about its previous actions.  For example, $p_j$ could increase, as the number of steps without food increases. This would provide a strategy in which the forager only executes long range movement if they are starving, which will probably limit the rate at which food is consumed and increase the overall lifetime $T$.  Our model could also incorporate greed (or anti-greed) as a parameter~\cite{bhat17}. As shown in previous work, the lifetime of foragers increases in one-dimensional environments if their diffusion is biased away from food. This finding mirrors our own conclusion, that foragers maximize their lifetime by balancing a reduction in the probability of early starvation with the conservation of resources.

\section*{\label{sec:acknowledgments}Acknowledgments}
NK was supported by EXTREEMS - QED: Directions in Data Discovery in Undergraduate Education (NSF DMS-1407340). ZPK was supported by an NSF grant (DMS-1615737).

\appendix
\section{Transition matrix for small system}
\label{appss}
We define the states of the small system with $n=4$ sites and survival time $s=2$ according to the relative location of the forager and the arrangement of food sites remaining. This numbers thirteen distinct states, with a transition matrix $Q$ for the update of the state vector $S_{t+1} = Q^T S_{t}$ where $Q_{1,2}=Q_{1,3}=Q_{2,3}=Q_{2,13}=Q_{9,10}=Q_{9,11}=Q_{10,11}=\frac{p_j}{4}$, $Q_{1,5}=Q_{2,5}=Q_{9,8}=\frac{2-p_j}{2}$, $Q_{3,4}=Q_{4,13}=Q_{7,8}=\frac{p_j}{2}$, $Q_{3,9}=Q_{4,9}=Q_{7,10}=Q_{7,11}=Q_{8,11}=\frac{2-p_j}{4}$, $Q_{5,6}=Q_{5,7}=Q_{6,7}=Q_{6,13}=\frac{1}{2}$, $Q_{8,13}=\frac{2+p_j}{4}$, $Q_{10,13}=\frac{4-p_j}{4}$, $Q_{11,12}=Q_{12,13}=Q_{13,13}=1$. Note that the thirteenth state is the absorbing cemetery state.

\end{document}